\documentclass[twocolumn,showpacs,preprintnumbers,amsmath,amssymb]{revtex4}

\usepackage{graphicx}
\usepackage{dcolumn}
\usepackage{bm}

\begin{document}

\preprint{}

\title{Formation of Deeply Bound Kaonic Atoms in ($K^-,N$) Reactions}

\author{J. Yamagata$^1$, H. Nagahiro$^2$, R. Kimura$^1$}
\author{S. Hirenzaki$^1$}%
\affiliation{$^1$Department of Physics, Nara Women's University, Nara 630-8506, Japan\\
$^2$Research Center for Nuclear Physics (RCNP), Osaka University, Ibaraki, Osaka 567-0047, Japan
}%

\date{\today}

\begin{abstract}
We study theoretically the ($K^-,N$) reactions for the formation of the deeply bound kaonic atoms, which were predicted to be quasi--stable with narrow widths, using the Green function method.
We consider various cases with different target nuclei and energies systematically and find the clear signals in the theoretical spectra for all cases considered in this article.
The signals show very interesting structures, such as the $RESONANCE~DIP$ instead of the resonance peak.
We discuss the origins of the interesting structures and possibilities to get new information on the existence of the kaonic nuclei from the spectra of the atomic state formations.
\end{abstract}

\maketitle

\section{\label{sec:intro}Introduction}
Kaonic atoms and kaonic nuclei 
carry important information concerning the $K^-$--nucleon interaction in nuclear medium.  This information is very important in understanding the kaon properties at finite density and, for example, to determine the constraints on kaon condensation in high--density matter.

In the beginning of this article, we need a few comments to clarify the differences of kaonic atoms and kaonic nuclei.
The kaonic nuclei are kaon--nucleus bound systems by way of the strong interactions inside the nucleus and are the controversial objects as described in Refs.~\cite{Akaishi02,finuda,oset06}.
We simply mention here that some indications of existence of the NARROW kaonic nuclear states reported in Refs.~\cite{Iwasaki03} were withdrawn by the authors in recent conference talks~\cite{iwasaki06}.
On the other hand, the kaonic atoms are the Coulomb assisted kaon--nucleus bound systems which were predicted to be quasi--stable theoretically by the realistic kaon--nucleus optical potentials.
Thus, the kaonic atoms are expected to be observed by the proper reactions.
Actually, the shallow atomic states have been observed experimentally by the kaonic X--ray spectroscopy for a long time.
In the beginning of Sec.~\ref{3}, we show two figures to explain the quantitative differences of kaonic atoms and kaonic nuclei.
We consider mainly the deeply bound kaonic atoms, which can not be observed by the standard X--ray spectroscopy, in this article.

We study the in--flight ($K^-,N$) reactions systematically using the Green function method~\cite{morimatsu85} for the formation of kaonic atoms in this article.
The Green function method is known to be suited for evaluations of formation rates both of stable and unstable bound systems. The ($K^-,N$) reaction for the ${\bar K}$--nucleus system formation was proposed in Refs.~\cite{Kishimoto99,Friedman99}, and theoretical results of the reaction energy spectra were obtained in Ref.~\cite{gata05} using the same theoretical approach of Ref.~\cite{hirenzaki91} for the deeply bound pionic atom formation reaction~\cite{tok,gilg00}, where we used the effective number approach, which is best suited for the studies of the discrete states formation. 
The theoretical spectra using the Green function method were later obtained in Ref.~\cite{gata06}, where we have shown the difficulties to obtain clear signals for kaonic nuclear states formation experimentally.
Similar results were also reported in Ref.~\cite{arima02}.
In our previous paper~\cite{gata06}, we also found simultaneously that there existed clear signals for the kaonic atom formations in the theoretical spectra which showed the interesting structures like $RESONANCE~~DIP$.
We think the evaluation obtained in Refs.~\cite{gata05,gata06} is very interesting and important to know the experimental feasibilities of the ($K^-,N$) reaction and to understand the deeper meanings of the observed spectra both for kaonic atoms and kaonic nuclei formations~\cite{Kishimoto99,Kishimoto03}.

The purpose of this article is to study thoroughly the atomic state formation spectra by ($K^-,N$) reactions.
We consider various cases with different target nuclei and incident energies and show the calculated spectra.
We then study the origins of the interesting structures of the spectra reported in this article and in Ref.~\cite{gata06}, and we study the experimental feasibilities.
We also discuss the possibilities to get new information on the kaon--nucleus interaction and the existence of the kaonic nuclei from the ($K^-,N$) spectra of the atomic state formation.
We believe that the realistic calculations of the formation spectra are necessary for all observed results to study the kaon properties in nuclear medium and to obtain the decisive conclusions.
So far, experimental studies of the in--flight ($K^-,N$) reactions were proposed for the studies of kaonic nuclei and performed by Kishimoto and his collaborators \cite{Kishimoto99,Kishimoto03}.
Our calculated results for kaonic nuclei formation~\cite{gata05,gata06} are in very good agreement with the data in Refs.~\cite{kishimoto_talk} reported after our predictions.

The $K^-$--nucleus interaction has been studied for a long time based on the shallow kaonic atom data obtained by the X--ray spectroscopy.
Very interesting feature of kaon--nucleus bound systems is based on the fact that the properties 
of kaons in nuclei are strongly influenced by the change undergone by 
$\Lambda(1405)$ in nuclear medium, because $\Lambda(1405)$ is a resonance 
state just below the kaon-nucleon threshold.  In fact, there are studies 
of kaonic atoms carried out by modifying the properties of $\Lambda(1405)$ in nuclear medium \cite{alberg76,wei,miz}. 
These works reproduce the properties of specific kaonic atoms reasonably well. In Ref.~\cite{batty97}, the phenomenological study of kaonic atoms are performed comprehensively, where the density--dependent potentials are considered for $\chi ^2$ fitting to take into account possible nonlinear effects that could be because of $\Lambda(1405)$ resonance.

Recently, there have been significant developments in the description
of hadron properties in terms of the $SU(3)$ chiral Lagrangian. The interpretation of the $\Lambda(1405)$ resonance state as a baryon--meson coupled system proposed in Ref.~\cite{dalitz67} is also supported by the studies with the chiral Lagrangian~\cite{kai,ose}. Subsequently, the properties of $\Lambda(1405)$
in nuclear medium 
using the $SU(3)$ chiral unitary model were also investigated by
Waas et al. \cite{waa}, Lutz \cite{lut}, Ramos and Oset \cite{ram}, and Ciepl$\acute{\rm y}$ et al. \cite{ciep01}. 
All of these works considered the Pauli effect on the 
intermediate nucleons. In addition, in Ref.~\cite{lut}, the self-energy of the 
kaon in the intermediate states is considered, and in Ref.~\cite{ram}, the 
self-energies of the pions and baryons are also taken into account.  These 
approaches lead to a kaon self-energy in nuclear medium that can be 
tested with kaonic atoms and kaonic nuclei. The in--medium ${\bar K}$ properties have been studied also in Refs.~\cite{tolo01,tolo02} based on meson--exchange J$\ddot{\rm u}$lich ${\bar K}N$ interaction, and in Ref.~\cite{schaffner00}, for the environment in heavy ion collisions. 
These theoretical potentials are shown to have the ability to reproduce the kaonic atom data reasonably well~\cite{hirenzaki00,baca00}.
The kaonic nuclear states were also studied using these interactions and shown to have large decay widths of the order of several tens of MeV.~\cite{hirenzaki00,gata05,gata06,gata07,friedman99,Friedman99}

In this work, we mainly use the phenomenological optical potential obtained in Ref.~\cite{batty97} and the microscopic chiral unitary optical potential obtained in Ref.~\cite{ram} to investigate the structure and formation of deeply bound kaonic atoms.
We also consider another type of phenomenological optical potential developed to take into account the two body kaon absorption effects explicitly according to Ref.~\cite{gata07}.

In Sec.~\ref{sec2}, we describe the theoretical models for the studies of the structure and formation of kaonic atoms. 
Numerical results are presented and discussed in Sec.~\ref{3}.
In Sec.~\ref{sec4}, we discuss the effects of existence of kaonic nuclei to the spectra in the energy region of the atomic states formation, and also discuss the origin of the interesting spectra shape of the atomic states formation. 
We give conclusions of this article in Sec.~\ref{5}.

\section{\label{sec2}Formalism}
To investigate the structure and formation of the kaonic atoms theoretically, we consider the Klein--Gordon equation
\begin{equation}
\label{KG_eq}
[-{\bm \nabla}^2+\mu^2+2\mu V_{\rm opt}(r)]\phi({\bm r})=[\omega-V_{\rm coul}(r)]^2 \phi({\bm r}).
\end{equation}

\noindent
Here, $\mu$ is the kaon--nucleus reduced mass and $V_{\rm coul}(r)$ is the Coulomb potential with a finite nuclear size:
\begin{equation}
\label{V_coul}
V_{\rm coul}(r)=-e^2 \int \frac{\rho_{\rm ch}(r')}{\bigl| {\bm r}-{\bm r'}\bigr|}d^3 r',
\end{equation}
\noindent
where $\rho_{\rm ch}(r)$ is the charge distribution of the nucleus. We employ the empirical Woods--Saxon form for the density as
\begin{equation}
\label{rho_ch}
\rho_{\rm ch}(r)=\frac{\rho_0}{1+\exp[(r-R)/a]},
\end{equation}
\noindent
where we use $R=1.18A^{1/3}-0.48$ fm and $a=0.5$ fm with $A$, the nuclear mass number. To evaluate the kaon--nucleus optical potential, we use the point nucleon density distributions deduced from the $\rho_{\rm ch}$ in Eq.~(\ref{rho_ch}) by the same prescription described in Sect.~4 in Ref.~\cite{nieves93}. The shapes of the density distributions of the proton and neutron are assumed to be the same in this article.
For the studies of the structure of the kaonic atoms, we solve the Klein--Gordon equation numerically, following the method of Oset and Salcedo~\cite{oset85}.
The application of this method to pionic atom studies is reported in detail in Ref.~\cite{nieves93}.

We use the Green function method~\cite{morimatsu85} to calculate the formation cross sections of the ${\bar K}$--nucleus system in the ($K^-,p$) reactions. The details of the application of the Green function method are found in Refs.~\cite{hayano99,klingl99,jido02,nagahiro05}. 

The present method starts with a separation of the reaction cross section into the nuclear response function $S(E)$ and the elementary cross section of the $p({\bar K},p){\bar K}$ with the impulse approximation
\begin{equation}
\label{crossS}
\Biglb( \frac{d^2\sigma}{d\Omega dE}\Bigrb)_{A({\bar K},p)(A-1)\otimes {\bar K}}=\Biglb( \frac{d\sigma}{d\Omega}\Bigrb)^{\rm lab}_{p({\bar K},p){\bar K}}\times S(E).
\end{equation}
\noindent
The forward differential cross section of the elementary process $p({\bar K},p){\bar K}$ in the laboratory frame $(d\sigma /d\Omega)^{\rm Lab}_{p({\bar K},p){\bar K}}$ is evaluated to be 8.8 mb/sr at $T_{\bar K}=600~{\rm MeV}$ using the $K^-p$ elastic cross--section data in Ref.~\cite{conforto76}. We should mention here that the corrections to this evaluation were reported in Ref.~\cite{ciep01}, which reduce the elementary cross section to be 3.6 mb/sr effectively. In this article, we show the all calculated results with assuming the elementary cross section to be 8.8 mb/sr.

The calculation of the nuclear response function with the complex potential is formulated by Morimatsu and Yazaki~\cite{morimatsu85} in a generic form as 
\begin{equation}
\label{S(E)}
S(E)=-\frac{1}{\pi} {\rm Im}\sum_f \tau^\dagger_f G(E) \tau_f,
\end{equation}
\noindent
where the summation is taken over all possible final states. The amplitude $\tau_f$ denotes the transition of the incident particle (${\bar K}$) to the proton--hole and the outgoing ejectile ($p$) , involving the proton--hole wave function $\psi_{j_p}$ and the distorted waves $\chi_i$ and $\chi_f$, of the projectile and ejectile, taking the appropriate spin sum
\begin{equation}
\label{tau}
\tau_f({\bm r})=\chi_f^*({\bm r})\xi_{1/2,m_s}^*[Y_{l_{\bar K}}^*(\hat{\bm r})\otimes \psi_{j_p}({\bm r})]_{JM}\chi_i({\bm r})~,~
\end{equation}
\noindent
with the meson angular wave function $Y_{l_{\bar K}}(\hat{\bm r})$ and the spin wave function $\xi_{1/2,m_s}$ of the ejectile. The distorted waves are written with the distortion factor $F({\bm r})$ as 
\begin{equation}
\label{chi}
\chi_f^*({\bm r})\chi_i({\bm r})=\exp(i{\bm q}\cdot{\bm r})F({\bm r})~,~
\end{equation}
\noindent
with the momentum transfer ${\bm q}$. The distortion factor $F({\bm r})$ is defined as
\begin{equation}
\label{distortionfactor}
F({\bm r})=\exp \Bigl( -\frac{1}{2}\bar{\sigma}\int^{\infty}_{-\infty}dz'\bar{\rho}(z',{\bm b})\Bigr)~,~
\end{equation}
where $\bar{\sigma}$ is the averaged distortion cross section defined as
\begin{equation}
\bar{\sigma}=\frac{\sigma_{{\bar K}N}+\sigma_{pN}}{2}~,
\end{equation}
with the total cross sections of the incident kaon and emitted proton with the nucleons in the nucleus.~The averaged nuclear density $\bar{\rho}(z',{\bm b}$) in Eq.~(\ref{distortionfactor}) is defined as
\begin{equation}
 \bar{\rho}(r)=\frac{\rho_0}{1+\exp[(r-\bar{R})/\bar{a}]}~
\end{equation}
in the polar coordinates with the averaged radial parameter $\bar{R}$ and diffuseness parameter $\bar{a}$ defined as
\begin{equation}
\bar{R}=\frac{R_i+R_f}{2}~
\end{equation}
and
\begin{equation}
\bar{a}=\frac{a_i+a_f}{2}~,
\end{equation}
with the density parameters of the nuclei in the initial and final states.\\

The Green function $G(E)$ contains the kaon--nucleus optical potential in the Hamiltonian $H_{\bar K}$ as,
\begin{equation}
\label{Green1}
G(E,{\bm r},{\bm r'})=\langle p^{-1}|\phi_{\bar K}({\bm r})\frac{1}{E-H_{\bar K}+i\epsilon}\phi^\dagger_{\bar K}({\bm r})|p^{-1}\rangle~,
\end{equation}
where $\phi^\dagger_{\bar K}$ is the meson creation operator, $|p^{-1}\rangle$ the proton--hole state, and $E$ the kaon energy defined as $E=T_{\bar K}-T_p-S_p$ using the kinematical variables defined in the formation reaction, where $T_{\bar K}$ is the incident kaon kinetic energy, $T_p$ the emitted proton kinetic energy, and $S_p$ the proton separation energy from the each proton single--particle level, which is compiled in Table~${\rm III}$ in Ref.~\cite{gata05} and in Table~\ref{tb:sp} in this article for various cases.
The separation energies for the ground states of the daughter nuclei can be found in ISOTOPE Table~\cite{isotope}.
Obtaining the Green function with the optical potential is essentially the same as solving the associated Klein--Gordon equation.~We can calculate the nuclear response function $S(E)$ from $\tau^\dagger_f({\bm r})G(E;{\bm r},{\bm r'})\tau_f({\bm r'})$ by performing appropriate numerical integrations for the variables ${\bm r}$ and ${\bm r'}$.

In the Green function formalism, we can calculate the response function $S(E)$ for both bound and quasifree kaon production energy regions, and we can also perform the summation of the kaon final states without assuming the existence of the discrete kaon bound states, which could disappear in the cases with the strongly absorptive optical potential.

As for the kaon--nucleus interaction, we mainly consider two different optical potentials: that obtained with the chiral unitary approach \cite{ram} and that obtained with a phenomenological fit \cite{batty97}.
We also consider another type of phenomenological potential in some cases for comparison, which was developed according to Ref.~\cite{gata07} to take into account the two body kaon absorption effect explicitly.
The optical potentials of the chiral unitary approach, which is obtained by the kaon self--energy in nuclear matter with the local density approximation, is described in detail in Ref.~\cite{ram}.

The optical potential obtained in the phenomenological fit~\cite{batty97} is written as
\begin{equation}
\label{V_opt_ph}
2\mu V_{\rm opt}=-4 \pi \eta a_{\rm eff}(\rho)\rho(r),
\end{equation}

\noindent
where $a_{\rm eff}(\rho)$ is a density--dependent effective scattering length and $\eta=1+m_{\bar K}/M_N$. The $a_{\rm eff}(\rho)$ is parameterized as
\begin{equation}
\label{real_aeff}
{\rm Re}~a_{\rm eff}=-0.15+1.66(\rho/\rho_0)^{0.24}~{\rm fm}
\end{equation}
and
\begin{equation}
\label{im_aeff}
{\rm Im}~a_{\rm eff}=0.62-0.04(\rho/\rho_0)^{0.24}~{\rm fm}.
\end{equation}

\noindent
When we consider the atomic states, the energy dependence of the potentials considered in Ref.~\cite{gata06} is irrelevant because of the relatively small binding energies of the atomic states.
We include the energy dependence of the phenomenological optical potentials in the exactly same manner in Refs.~\cite{gata06,mares04} whenever it is necessary.

Another type of the phenomenological optical potential has the same real part as that of the phenomenological potential described above (Eq.~(\ref{real_aeff})).
The imaginary part of the potential has different form to show the two body kaon absorption effect explicitly as explained in Ref.~\cite{gata07} in detail,
\begin{equation}
{\rm Im}V_{\rm opt}=-\frac{4\pi}{2\mu}\bigl{\{}\bigl( 1+\frac{m_{\bar K}}{m_N}\bigr) a_1\rho+\bigl( 1+\frac{m_{\bar K}}{2m_N}\bigr) a_2 \rho^2\bigr{\}}~~.
\label{KYH}
\end{equation}
Two parameter sets are newly determined to obtain better $\chi^2$ fitting to kaonic atom data for various nuclei than the parameters listed in Ref.~\cite{gata07}, where we only consider $K^--^{12}$C system.
One set is $(a_1,a_2)=(1.15,66.8)$ in kaon mass unit which are determined by assuming the kaon absorption from atomic $2p$ states~\cite{gata07}, and another set is $(a_1,a_2)=(0.91,136)$ in kaon mass unit determined by assuming the kaon absorption from atomic $3d$ states~\cite{gata07}.
In this article, we adopted the potential parameter $(a_1,a_2)=(1.15,66.8)$.
We use the same energy dependence of the potential in Ref.~\cite{gata07} whenever it is necessary.

\section{\label{3}Numerical Results}
\begin{figure}[htpd]
\includegraphics[width=8cm,height=6cm]{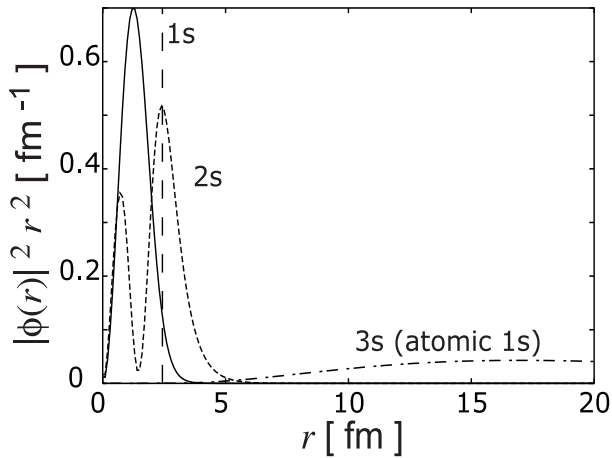}
\caption{\label{fig:wavefunc}The kaonic bound state density distributions $|\phi (r)|^2r^2$ in coordinate space for $^{15}$N obtained with the phenomenological optical potential~\cite{batty97}.
The solid and dotted curves indicate the distributions of $1s$ and $2s$ states.
The dashed curve represents the density of the $3s$ state which is usually regarded as a kaonic atom $1s$ state.
The half--density radius of $^{15}$N is also shown.}
\end{figure}

\begin{figure}[htpd]
\includegraphics[width=8cm,height=6cm]{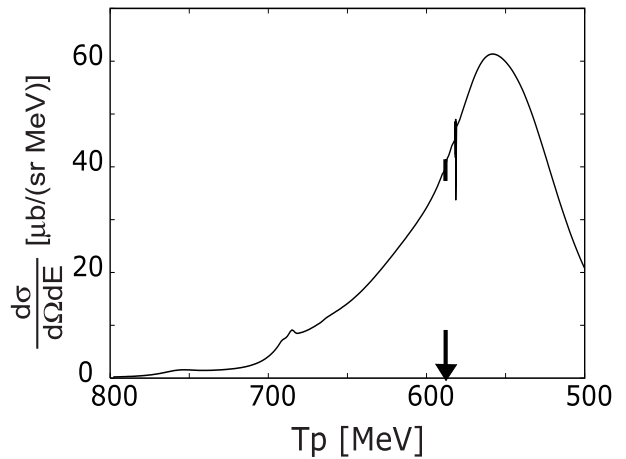}
\caption{\label{fig:16Ospectrum}
Calculated spectra of the $^{16}$O($K^-,p$)$^{15}$N$\otimes K^-$ reactions at $T_{\bar K}=600$~MeV plotted as a function of the emitted proton energy $T_p$ at $\theta^{\rm lab}_p=0$~(deg.).
The energy dependent phenomenological potential is used as in Ref.~\cite{gata06}.
The arrow indicates the kaon production threshold.}
\end{figure}

First, we show the figures of the calculated results of the density distributions of kaon bound states (Fig.~\ref{fig:wavefunc}), and of the ($K^-,p$) reaction spectra (Fig.~\ref{fig:16Ospectrum}) for the kaonic atoms and kaonic nuclei formation of the $K^-$--$^{15}$N system to explain the differences of the kaonic atoms and kaonic nuclei.
The both states are the solutions of the Klein--Gordon equation Eq.~(\ref{KG_eq}) with the same optical potential.
Thus, they should be called sequentially as $1s$, $2s$, $3s~~\cdots$ for $s$--wave states for example.
However, since the real part of the optical potential is strongly attractive, there exist kaonic nuclear states bound in the interior of the nucleus which can be easily distinguished from the atomic `outer' states in the series of eigen states of the same Hamiltonian as we can see in Fig.~\ref{fig:wavefunc}.
 In Fig.~\ref{fig:wavefunc}, we show the density distributions of the $s$--wave kaonic bound states in $^{15}$N.
Both the ground state ($1s$) and the first excited state ($2s$) exist almost inside the nucleus, while the second excited state ($3s$) behaves like the ground state in Coulomb potential which usually recognized as the `ground state of kaonic ATOM'.
As we can expect from the density distributions, the `interior' states ($1s$ and $2s$) and the `outer' state ($3s$) must have much different properties.
Thus, we call the interior states as the kaonic nuclei and the outer states as the kaonic atoms to distinguish these states with significantly different characters.
This classification can be easily and unambiguously performed by observing the densities, for example, as shown in Fig.~\ref{fig:wavefunc}.

In Fig.~\ref{fig:16Ospectrum}, we show the calculated ($K^-,p$) reaction spectra for the formation of $K^-$--$^{15}$N bound systems.
Here, we can also see the differences of the atomic states and nuclear states.
In the spectra, the contributions of the formation of nuclear states distribute in the wide energy range like $T_p=$600--800~MeV and do not make any clear signals because of the large decay widths in our calculation.
On the other hand, the atomic states have smaller binding energies and widths, and make clear signals around the threshold $T_p\sim 590$~MeV shown as two spikes in Fig.~\ref{fig:16Ospectrum}.
The detail structures of the spikes will be explained later.
In this article, we mainly consider the structure and formation of these atomic `outer' states. 

\begin{table}[htbp]
\begin{center}
\caption{Calculated binding energies and widths of kaonic atoms in $^{119}$In and $^{207}$Ti with the phenomenological optical potential~\cite{batty97} 
in unit of keV.}
\begin{tabular}{c|cc|cc}
\hline
\hline
&&&&\\
Atomic State& \multicolumn{2}{c|}{$^{119}$In}&\multicolumn{2}{|c}{$^{207}$Ti}\\ 
(keV)&B.E.& $\Gamma$&B.E.& $\Gamma$\\ 
\hline
1$s$&~4309.7~&~999.7~&~7279.9~&~1544.3\\
2$s$&~2226.6~&~391.0~&~4194.2~&~711.5\\
3$s$&~1376.8~&~194.7~&~2775.7~&~393.7\\
4$s$&~938.2~&~111.0~&~1982.7~&~241.6\\
2$p$&~3774.2~&~740.2~&~6779.2~&~1335.6\\
3$p$&~2015.2~&~304.7~&~3962.3~&~629.2\\
4$p$&~1270.5~&~156.4~&~2646.7~&~353.1\\
3$d$&~2890.7~&~328.2~&~5853.3~&~896.4\\
4$d$&~1658.7~&~160.2~&~3536.3~&~451.9\\
4$f$&~1930.7~&~54.2~&~4629.0~&~434.4\\
\hline
\end{tabular}
\label{tab:ph}
\end{center}
\end{table}

\begin{table}[htbp]
\begin{center}
\caption{Calculated binding energies and widths of kaonic atoms in $^{119}$In and $^{207}$Ti with the optical potential of the chiral unitary model~\cite{ram} 
in unit of keV.}
\begin{tabular}{c|cc|cc}
\hline
\hline
&&&&\\
Atomic State& \multicolumn{2}{c|}{$^{119}$In}&\multicolumn{2}{|c}{$^{207}$Ti}\\ 
(keV)&B.E.& $\Gamma$&B.E.& $\Gamma$\\ 
\hline
1$s$&~4296.8~&~944.5~&~7241.8~&~1502.7\\
2$s$&~2226.2~&~366.7~&~4186.5~&~676.6\\
3$s$&~1377.5~&~182.2~&~2774.5~&~372.0\\
4$s$&~938.9~&~103.7~&~1983.1~&~227.8\\
2$p$&~3778.4~&~681.0~&~6755.9~&~1270.8\\
3$p$&~2020.5~&~279.9~&~3960.2~&~586.8\\
4$p$&~1273.9~&~143,4~&~2648.5~&~328.1\\
3$d$&~2888.2~&~303.8~&~5837.0~&~858.2\\
4$d$&~1658.9~&~147.6~&~3532.9~&~424.0\\
4$f$&~1930.9~&~48.4~&~4626.0~&~404.8\\
\hline
\end{tabular}
\label{tab:chi}
\end{center}
\end{table}

\begin{figure}[htpd]
\includegraphics[width=9cm,height=5cm]{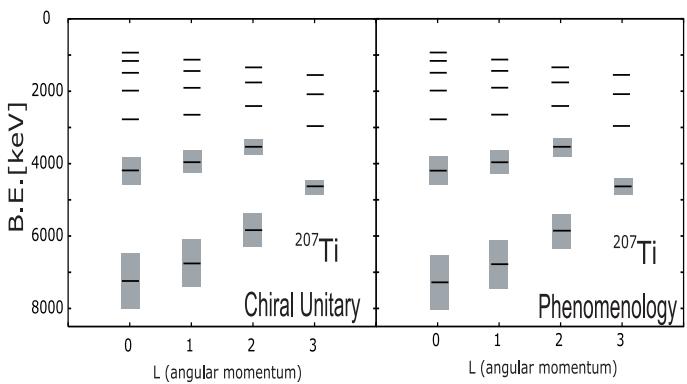}
\caption{\label{fig:level}
Energy levels of kaonic atoms of $^{207}$Ti up to the principle quantum number $n=7$ obtained with the optical 
potentials of the chiral unitary model (left) and of the phenomenological fit (right). 
The hatched areas indicate the level widths for the deeply bound states such as B.E. $>$ 3 MeV. 
}
\end{figure}

We solve the Klein--Gordon equation with the optical potential shown in Eq.~(\ref{KG_eq}) to obtain the binding energies and widths of the atomic states.
The results for lighter nuclei were already compiled in Tables~I and ${\rm II}$ of Ref.~\cite{gata05}. We show here the calculated results for heavier nuclei, $^{119}$In and $^{207}$Ti, which are corresponding to the kaon--nucleus systems formulated by the $^{120}$Sn($K^-,p$) and $^{208}$Pb($K^-,p$) reactions.
We show the calculated binding energies and widths of kaonic atoms in Table~\ref{tab:ph} and \ref{tab:chi} for the phenomenological optical potential~\cite{batty97} and the chiral unitary optical potential~\cite{ram}, respectively.
We also show the level structure of the kaonic atoms in $^{207}$Ti in Fig.~\ref{fig:level}.
As we found before in our previous paper~\cite{gata05}, the results obtained by the both potentials resemble each other and, at the same time, indicate the existence of the discrete deeply bound atomic states in all nuclei generally.
For example, as shown in Fig.~\ref{fig:level}, the atomic $1s$, $2p$, $3d$ states overlap each other due to the level widths.
However, we can also expect to have other narrow atomic states.
Actually we should notice here that the kaonic X--ray spectroscopy only provides the data for $7i$ state in this nuclear mass region.
So, deeper states than the $7i$ state have not been observed so far, and should be studied experimentally.
As we mentioned above, the theoretical predictions are quite robust in this binding energy region and are quite reliable.
The similar results for the atomic level structure with the phenomenological potential were also reported in Ref.~\cite{friedman99}.

To evaluate the formation spectra of these atomic states by the ($K^-,N$) reactions, we use the Green function method.
As we have explained in Sec.~\ref{sec2}, in the ($K^-,N$) reactions one nucleon is picked--up and emitted from the target nucleus in the final states.
We consider here the ($K^-,p$) reactions mainly.
The deeper proton--hole states have large decay width for $\gamma$--decay and are not suited for the kaonic atom formation.
Because the widths of the proton--hole states can be significant larger than the level spacing of kaonic atoms and can smear out all signals.
Thus, we consider the only proton single particle levels corresponding to the ground states of the daughter nuclei which do not have decay widths.
We show in Table~\ref{tb:sp} the proton single particle levels ($j_p$) and the separation energies ($S_p$) used in this paper, which are taken from Ref.~\cite{isotope}.
\begin{table}[htbp]
\begin{center}
\caption{One proton separation energies $S_p$ of $^{12}$C, $^{16}$O, $^{40}$Ca, $^{120}$Sn, and $^{208}$Pb take from Ref.~\cite{isotope}.
All proton--holes in the final states are corresponding to the ground states of the daughter nuclei and have no decay widths.}
\vspace{3mm}
\begin{tabular}{|c|c|c|} 
\hline
&single particle states&\\
~Target~&$~~j_p~~$&$~~S_p$~[MeV]~~\\
\hline
$^{12}$C&$~~~1p_{3/2}~~$&16.0\\
$^{16}$O&$1p_{1/2}$&12.1\\
$^{40}$Ca&$1d_{3/2}$&8.3\\
$^{120}$Sn&$1g_{9/2}$&10.7\\
~~$^{208}$Pb~~&$3s_{1/2}$&8.0\\
\hline
\end{tabular}
\label{tb:sp}
\end{center}
\end{table}

\begin{figure}[htpd]
\includegraphics[width=7.5cm,height=6cm]{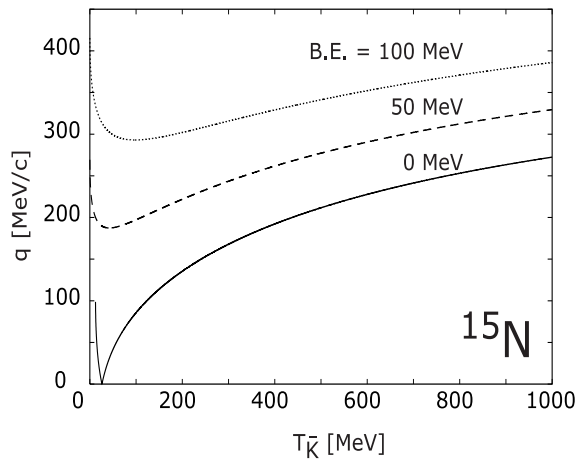}
\caption{\label{fig:momentum}
Momentum transfer in the $^{16}$O($K^-,p$)$^{15}$N$\otimes K^-$ reactions.
The proton separation energy $S_p$ is fixed at 12.1 MeV, and the kaon binding energies are assumed to be 0 MeV (solid curve), 50 MeV (dashed curve), and 100 MeV (dotted curve). }
\end{figure}

As you can see in the studies of the deeply bound $\pi$ atom formation reactions~\cite{hirenzaki91,gilg00}, the matching condition of the momentum transfer and the angular momentum transfer plays an important role to provide the large formation rate.
As we can see from Table~\ref{tb:sp}, the target nuclei considered in this paper have various proton--hole states with different quantum numbers and we can expect to have the interesting spectra with the characteristic behaviors for different targets.
In Fig.~\ref{fig:momentum}, we show the momentum transfer of the $^{16}$O($K^-,p$)$^{15}$N$\otimes K^-$ reactions as the functions of the incident kaon kinetic energy for several kaon binding energy cases.
We find that the recoilless condition is only satisfied for kaon bound states with relatively small binding energies (B.E. $\leq$ several MeV), for small incident kaon energies.
For the states with large B.E. like kaonic nuclear states, the recoilless condition is never satisfied and, thus, the finite angular momentum transfer is preferred by the matching condition.
For kaonic atoms, we can expect to have nearly recoilless kinematics for the small incident kaon energies. On the other hand, the momentum transfer is expected to be $q\sim 200$ MeV/c for the kaon energies at J--PARC facility.

\begin{figure*}[htpd]
\includegraphics[width=12cm,height=7cm]{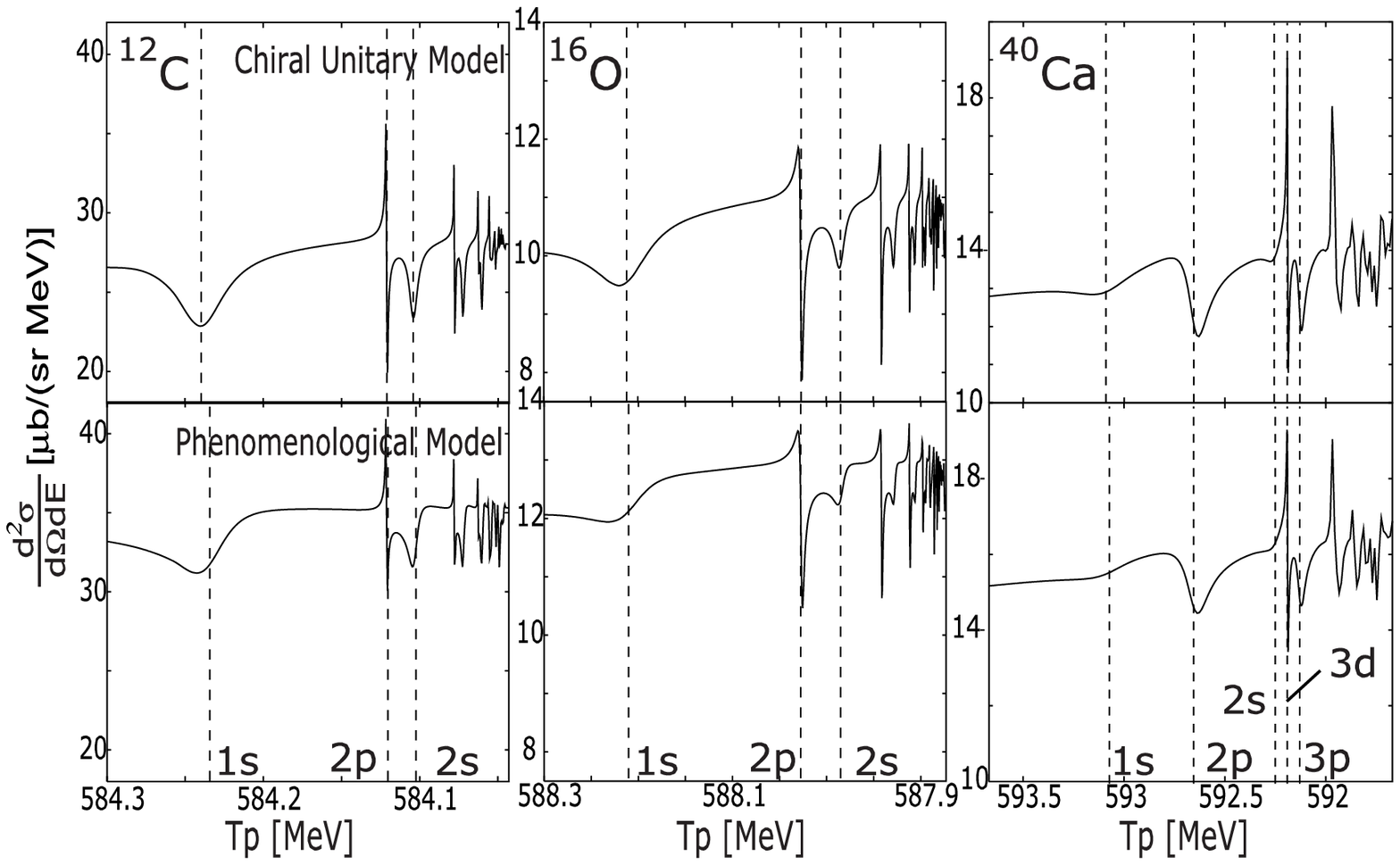}
\caption{\label{fig:atom}
Kaonic atom formation cross sections in ($K^-,p$) reactions at $T_{\bar K}=600$ MeV for $^{12}$C (left), $^{16}$O (center), and $^{40}$Ca (right) target nuclei plotted as functions of the emitted proton energy $T_p$ at $\theta_p^{\rm lab}=0$ (deg.) for the chiral unitary potential (upper)~\cite{ram} and for the phenomenological potential (lower)~\cite{batty97}.
The vertical dashed lines indicate the biding energies of some kaonic atom states as indicated in the figure.
Proton single particle state considered for each target is $^{12}$C ($1p_{3/2}$), $^{16}$O ($1p_{1/2}$), and $^{40}$Ca ($1d_{3/2}$) as listed in Table~\ref{tb:sp}.}
\end{figure*}

We show the calculated results of the kaonic atom formation spectra by in--flight kaon beam for light nuclei in Fig.~\ref{fig:atom}.
As reported in Ref.~\cite{gata06}, we find very interesting shapes of the spectra for all nuclei.
The calculated result for $^{12}$C target with the chiral unitary optical potential (left--upper figure in Fig.~\ref{fig:atom}) is essentially the same with that in Fig.~5 in Ref.~\cite{gata06}.
In this article, since we only include one proton--hole state in the final states corresponding to the ground states of the daughter nuclei, the absolute value of the cross sections are smaller than previous results~\cite{gata06}.

As we can see from Fig.~\ref{fig:atom}, the interesting structure of the spectra could be characterized by the angular momenta of the proton-- and kaon--single particle states.
For $^{12}$C target case, the proton--hole state is considered to be $1p_{3/2}$.
In this case, the subcomponents coupled with the $l_{\bar K}=0$ states ($1s, 2s, \cdots$) make the dips, while the subcomponents coupled with the $l_{\bar K}=1$ states ($2p, 3p, \cdots$) make the spectral shape with the steep rise and fall putting the resonance energy in between.
We can see the qualitatively same characteristic shapes for the $^{16}$O target cases in Fig.~\ref{fig:atom}, where the proton hole state is the $1p_{1/2}$ state.
On the other hand, the spectral shape has different features for the cases of $^{40}$Ca target with the $1d_{3/2}$ proton--hole state in the daughter nucleus.
In this case, the $l_{\bar K}=1$ states show the dip structure and the $l_{\bar K}=2$ states has the spectral shape with the steep rise and fall.
As for the origin of these interesting spectral shape, we will add some discussions in Sec.~\ref{sec4_2}.

In Fig.~\ref{fig:atom}, we can also see that the formation spectra calculated with the phenomenological optical potential and the chiral unitary potential are qualitatively same.
Thus, we can expect that the theoretical predictions are quite robust for these potentials.
On the other hand, to distinguish the different potentials, we need very precise data which enable us to recognize the tiny differences appeared between the upper and lower figures in Fig.~\ref{fig:atom}.

\begin{figure*}[htpd]
\includegraphics[width=11cm,height=8cm]{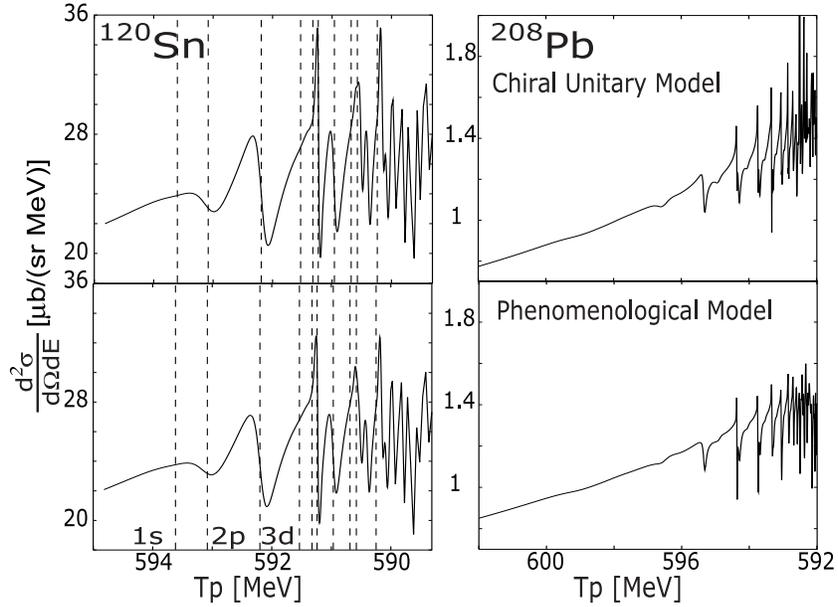}
\caption{\label{fig:atom2}
Kaonic atom formation cross sections in ($K^-,p$) reactions at $T_{\bar K}=600$ MeV for $^{120}$Sn (left) and $^{208}$Pb (right) target nuclei plotted as functions of the emitted proton energy $T_p$ at $\theta_p^{\rm lab}=0$ (deg.) for the chiral unitary potential (upper)~\cite{ram} and for the phenomenological potential (lower)~\cite{batty97}.
The vertical dashed lines indicate the binding energies of some kaonic atom states for $^{120}$Sn as indicated in the figure.
For $^{208}$Pb target (right), the bound states up to $4d$ do not show the clear indications in the spectra.
Proton single particle state considered for each target is $^{120}$Sn ($1g_{9/2}$) and $^{208}$Pb ($3s_{1/2}$) as listed in Table~\ref{tb:sp}.}
\end{figure*}

In Fig.~\ref{fig:atom2}, we also consider the heavier target cases, $^{120}$Sn and $^{208}$Pb.
The purposes to consider the heavier targets are to investigate the spectral shapes with different proton single particle levels, $1g_{9/2}$ for $^{120}$Sn and $3s_{1/2}$ for $^{208}$Pb, and to study the possibilities to observe very deeply bound kaonic atoms.
Actually, for the heavier nuclei the atomic level spacing is expected to be larger in general than the lighter nuclei cases due to the Coulomb like level structure, and thus, the very high energy resolution could be unnecessary to observe the peaks in the formation spectra and to distinguish the results with both potentials.

As we can see in Fig.~\ref{fig:atom2}, the level spacing and the width of each peak structure are larger for heavier targets cases than those for lighter targets shown in Fig.~\ref{fig:atom} as we expected.
The spectrum for $^{120}$Sn target case, where the proton $1g_{9/2}$ state is considered, show the qualitatively same structure with that for $^{40}$Ca target with $1d_{3/2}$ proton state , and has the dips for $l_{\bar K}=1$ states and the shape with the rise and fall for $l_{\bar K}=2$.
For $^{208}$Pb target case, we can only observe the clear signals for the formation of shallow atomic states at $T_{\bar K}=600$ MeV.
The bound states up to $4d$ do not show the clear indications due to the large widths and the large momentum transfer.
In both target cases, the shapes of the calculated spectra of both potentials, chiral unitary model and phenomenological model, resemble each other and it seems difficult to distinguish these potentials by the observed spectra.

\begin{figure}[htpd]
\includegraphics[width=5.5cm,height=9.0cm]{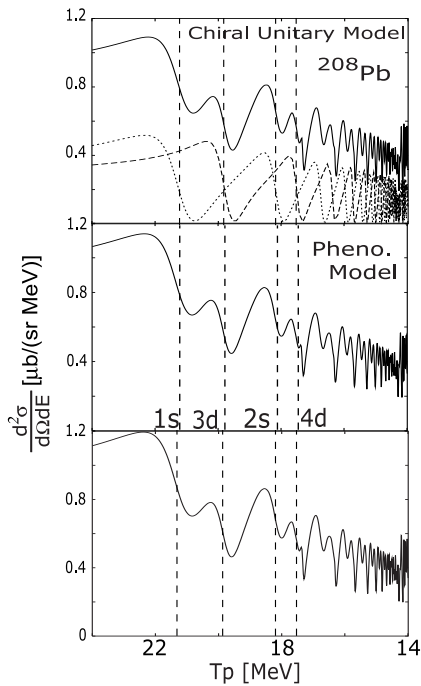}
\caption{\label{fig7}
Kaonic atom formation cross sections in $^{208}$Pb($K^-,p$) reactions at $T_{\bar K}=22$ MeV, which corresponds to the recoilless kinematics for the $^{208}$Pb target, plotted as the functions of the emitted proton energy $T_p$ at $\theta^{\rm lab}_p=0$ (deg.) for the chiral unitary potential (upper)~\cite{ram}, for the phenomenological potential (middle)~\cite{batty97}, and another phenomenological potential defined in Eq.~(\ref{KYH}) with the parameters, ($a_1,a_2$)=(1.15,66.8) in kaon mass unit (lower).
The dotted and dashed curves in the upper figure indicate the subcomponents of $s$ and $d$ wave kaonic atom states, respectively, and the vertical dashed lines indicate the binding energies of $s$ and $d$ wave kaonic atom states.
Proton single particle state considered is $3s_{1/2}$ as listed in Table~\ref{tb:sp}.
}
\end{figure}

Finally, we show the calculated results for the lower kaonic incident energy cases to know the spectra with the recoilless condition.
In Fig.~\ref{fig7}, we show the calculated spectra for $^{208}$Pb target cases.
Here, as shown in Eqs.~(\ref{chi}) and (\ref{distortionfactor}), we have not included Coulomb distortion for the emitted proton in our calculation, which could reduce the absolute cross sections for the lower proton energy cases.
In the recoilless kinematics, the kaonic states with small angular momenta are expected to be largely populated with the $s_{1/2}^{-1}$ proton hole for the $^{208}$Pb target because of the matching condition.
Actually, the $s$ and $d$ kaonic state contributions dominate the spectra.
The kaonic $p$ states are suppressed by the parity.
We have found that the calculated spectra with the recoilless condition shown in Fig.~\ref{fig7} are significantly different from those with the in--flight kaon cases shown in Fig.~\ref{fig:atom2} (right) and have found that the deeper kaonic states with lower angular momenta provide the clear structures in the spectra.
Hence, we think that the atomic states formation with the recoilless kinematics using the low energy kaon is also interesting to study the deeply bound kaonic atoms.
On the other hand, as shown in Fig.~\ref{fig7}, we have found again that the differences of the expected spectra due to those of the kaon--nucleus optical potential are tiny at the recoilless kinematics case.

\section{\label{sec4}Discussions}
\subsection{\label{sec4_1}Effects of Existence of Kaonic Nuclei}
In this section, we would like to explore the possible effects to the ($K^-,p$) spectra in the energy region of the atomic states formation due to the existence of the kaonic nuclear states.
As we discussed in the beginning of the Sec.~\ref{3}, the kaonic nuclear states and kaonic atom states are the eigen states of the same Hamiltonian, and their wave functions are expected to satisfy the orthogonal condition.$^{\footnotemark[1]}$
\footnotetext[1]{ Strictly speaking, this orthogonal condition holds only approximately for the Klein--Gordon equation Eq.~(\ref{KG_eq}) and should be modified from its usual form due to the existence of the energy dependence and the imaginary part of the optical potential \cite{nose97}.}
Thus, if there exist kaonic nuclear states, the wavefunctions of the kaonic atom states have nodes inside the nucleus to satisfy the orthogonality conditions with kaonic nuclear states.

On the other hand, the spatial dimensions of the kaonic nuclear states are almost equivalent to those of nucleons in nucleus.
Hence, we can expect that the wavefunctions of nucleons and kaonic nuclear states with the same quantum numbers are resemble each other, which means that the nucleon wavefunctions may satisfy the orthogonality conditions approximately with outer kaonic atom wavefunctions.

For example, we consider $K^--^{39}$K system with the phenomenological optical potential, which has the several kaonic nuclear states $1s, 2s, 2p, \cdots$ as listed in the Table ${\rm I}$ in Ref.~\cite{gata05}.
Thus, the kaonic atom wavefunctions of the system are expected to be approximately orthogonal to kaonic nuclear wavefunctions and also to the nuclear wavefunctions with the same principal quantum number ($n$) and orbital angular momentum ($l$).

Another important thing is that the strength of the formation of kaonic atoms can be simply estimated by the effective number approach adopted in Ref.~\cite{gata05,hirenzaki91}, which can be roughly described as,
\begin{equation}
N_{\rm eff} \propto |\int d{\bm r}\exp(i{\bm q}\cdot{\bm r})D({\bm b})\phi_{\bar K}^*\psi_N|^2
\label{neff1}
\end{equation} 
with the momentum transfer ${\bm q}$, the distortion factor $D({\bm b})$ and the wavefunctions of kaon and nucleon $\phi_{\bar K}^*$ and $\psi_N$~\cite{gata05,hirenzaki91}.
This factor can be reduced to the simpler form
\begin{equation}
N_{\rm eff}\propto |\int d{\bm r}\phi_{\bar K}^*\psi_N|^2~~,
\label{neff2}
\end{equation}
by considering the recoilless kinematics (${\bf q}\sim 0$) and neglecting the distortion effect, which will be $0$ for the orthogonal sets of the wavefunctions.
Thus, we can conclude that the existence of the kaonic nuclear states may significantly reduce the formation rate of kaonic atom stats with coupled to the nucleon hole states with the same quantum ($nl$) numbers of the kaonic nuclei in the recoilless kinematics.

We show the numerical results of these scenarios for some cases below.
We consider the ($K^-,n$) reactions for $^{40}$Ca and $^{16}$O targets.
The valence neutron states are 2$s_{1/2}$ for $^{40}$Ca and 1$p_{1/2}$ for $^{16}$O, respectively.
As shown in Table ${\rm I}$ and ${\rm II}$ in Ref.~\cite{gata05}, we expect to have kaonic nuclear $2s$ state for $A\sim 40$ nuclei with the phenomenological deep optical potential, while we do not expect the $2s$ kaonic states with the chiral unitary optical potential.
Similarly, we expect kaonic nuclei $p$ state for $A\sim 16$ nuclei with the phenomenological potential but not with the chiral unitary potential.
Thus, as described above, we can expect the significant suppression of the atomic state formation cross sections for the calculations with the phenomenological optical potential because of the approximate orthogonality between kaonic atom wavefunctions and valence neutron wavefunctions at the recoilless kinematics.  

\begin{figure}[htpd]
\includegraphics[width=7.5cm,height=8cm]{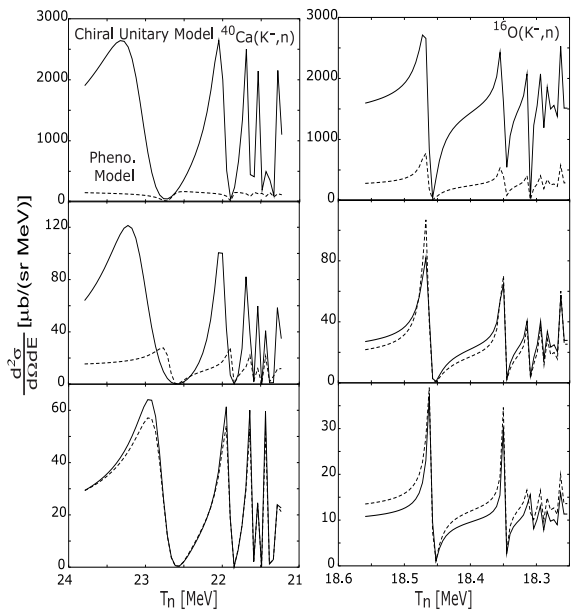}
\caption{\label{fig8}
Kaonic atom formation cross sections in $^{40}$Ca($K^-,n$) (left) and $^{16}$O($K^-,n$) (right) reactions at recoilless kinematics using low energy kaon beams with $T_{\bar K}=39.4$~MeV for $^{40}$Ca target and $T_{\bar K}=33.9$~MeV for $^{16}$O target.
The cross sections are plotted as the functions of the emitted neutron energy $T_n$ at $\theta_n^{\rm Lab}=0$ (deg.) for the chiral unitary potential (solid line)~\cite{ram} and the phenomenological potential (dashed line)~\cite{batty97}.
The spectra are calculated with (upper panels) no distortion effects ($F(\bf{r}$)=1~in Eq.~(\ref{chi})) and the reduced imaginary potentials Im$V$/5, (middle panels) full distortion effects and the reduced imaginary potentials Im$V$/5, and (lower panels) full distortion effects and full imaginary potentials corresponding to the full calculation in our model.
The neutron separation energies $S_n$ are 18.1 MeV for $^{40}$Ca and 15.7 MeV for $^{16}$O.}
\end{figure}

The numerical results are shown in Fig.~\ref{fig8}.
In the upper panels, we calculate the spectra by neglecting the distortion effects and by reducing the strength of the imaginary part of the optical potentials to Im$V_{\rm opt}$/5 to confirm our expectation in the ideal cases.
We have found that our expectation is correct and the spectra with the phenomenological optical potential is suppressed significantly in both target cases.
The spectra with the chiral unitary potential, which do not provide the kaonic nuclear states with the same quantum number as the valence neutron states, have significantly larger absolute values than those with the phenomenological potential as expected.
Now, we move to the realistic calculations.
In middle panels, we show the spectra with the full distortion effects and the reduced imaginary optical potentials Im$V_{\rm opt}$/5.
We found that the significant differences of the spectra due to the optical potential remains for $^{40}$Ca target case, while the differences disappeared for the $^{16}$O target case because of the distortion effects.

In the lower panels, we show the results with the full calculation of our model with the full distortion effects and the full imaginary part of the optical potentials.
We found that the discrepancies of the spectra for $^{40}$Ca target also disappeared by including the full imaginary part of the optical potentials, and the spectra with the different optical potentials resemble each other for both target cases.

Hence, we think that our idea is nice and reasonable, however, it doesn't work well for kaon--nucleus systems because of the strong absorption effects and the distortion effects of the reaction.
Thus, our idea could be realized in the cases with less absorptive systems formation by the reactions with the particles with smaller distortion effects.

\subsection{\label{sec4_2}Origin of interesting spectra Shape of atomic state formations}
In this section, we consider the origin of the interesting structures of the formation spectra of the kaonic atom states.
We follow the discussions given in Ref.~\cite{morimatsu85} and show a simple example.

\begin{figure}[htpd]
\includegraphics[width=6.5cm,height=5cm]{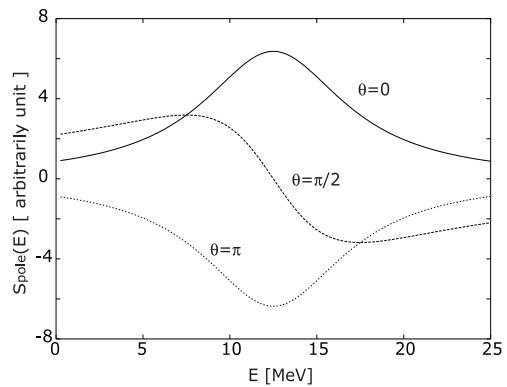}
\caption{\label{fig9}
The contribution to the spectra from the pole $S_{\rm pole}$ is plotted as functions of $E$.
The pole energy is assumed to be $E=12.5-5i$ MeV.
These lines indicate the $S_{\rm pole}$ behavior with the arg$A=0, \pi/2, {\rm and~} \pi$, respectively, where $A$ is defined in Eq.~(\ref{func:Atau}) in the text.}
\end{figure}

In the Green function formalism, we can divide the Green function into two parts, the pole part $G_{\rm pole}$ and the other background part $G_{\rm BG}$, as
\begin{equation}
G(E)=G_{\rm pole}(E)+G_{\rm BG}(E)~~,
\label{funcG}
\end{equation}
where the pole part can be written as,
\begin{equation}
G_{\rm pole}=\displaystyle{\frac{1}{E-(\epsilon-i\Gamma/2)}}\Bigl[{\rm res}G(E)\Bigl{|}_{E=\epsilon-i\Gamma/2}\Bigr]~~,
\label{func:Gpole}
\end{equation}
where $\epsilon-i\Gamma/2$ is the complex pole energy.
The contribution to the spectra from the the pole part can be expressed as
\begin{equation}
S_{\rm pole}(E)=-\displaystyle{\frac{1}{\pi}}{\rm Im}\displaystyle{\frac{A}{E-(\epsilon-i\Gamma/2)}}~~,
\label{func:Spole}
\end{equation}
where $A$ is defined as
\begin{equation}
A=\sum_f\tau^+_f\Bigl[{\rm res}G(E)\Bigl|_{E=\epsilon-i\Gamma/2}\Bigr]\tau_f
\label{func:Atau}
\end{equation}
with $\tau_f$ defined in Eq.~(\ref{tau}).
The $A$ includes the information of the residue of the Green function and the wavefunctions of the particles participating in the formation reactions.
As a simple example, we write the argument of the $A$ as $\theta={\rm arg}A$ and plot the $S_{\rm pole}$ in Eq.~(\ref{func:Spole}) for three $\theta$ values in Fig.~\ref{fig9}.

We have found that the shape of the formation spectra of the bound states changes according to the argument $\theta$ of the $A$, and we have the resonance peak structure for $\theta=0$, the spike structure with rise and down of the spectra for $\theta=\pi/2$, and the resonance dip for $\theta=\pi$.

Thus, we can deduce the information of the phase of the $A$ in Eq.~(\ref{func:Atau}) from the shape of the spectra.
On the other hand, the simple fit by the Lorentz distributions is not suited for these spectra and we should be careful to extract the resonance energies and the widths from the data.

\section{\label{5}Conclusion}
We study the formation spectra of the deeply bound kaonic atoms theoretically in this paper.
The deeply bound atomic states can not be observed by the standard X--ray spectroscopy, however, theoretical calculations have shown the existence of the narrow (quasi--stable) bound states beyond the accessible levels by the X--ray method.
We show the calculated results of the binding energies and widths of the deeply bound kaonic atom states with the phenomenological and the chiral unitary optical potentials.

As the formation reactions of the deeply bound states, we consider the missing mass spectroscopy by the ($K^-,N$) reactions.
We consider the various target nuclei and the different incident energies, and show the expected ($K^-,p$) spectra systematically.
We also consider ($K^-,n$) reactions in several cases.
And we also compare the expected spectra of the different optical potentials obtained by the phenomenological fit and the chiral unitary model.
We have found that the expected spectra always show the interesting structures for various target cases with the clear signals of the atomic states formation.

The origin of the interesting structure is also considered.
We show that the argument of the pole contribution of the spectral function is important and we should be careful to deduce the resonance energies and widths from the data.

The effects due to the existence of the kaonic nuclear states are also investigated.
We show that the exotic nuclear states with less absorptive interaction will be identified by observing the atomic states by the missing mass spectroscopy using the transport particles at the recoilless kinematics.

Finally, we would like to make a few comments on the importance of the experimental investigations of this research.
As shown in the present article, the structures of the deeply bound kaonic atoms are very robust and all optical potentials, which reproduce the existing shallow atomic data, are expected to provide the similar predictions for the deeply bound atomic states.
Thus, to confirm the predicted structures by the experiments is of great importance because the all optical potentials could be denied by the one experiment which proves no--existence of the deeply bound atoms.
It could happen if we all miss the important ingredients of the kaon--nucleus interactions for higher densities.
Thus, we think that it is very important to investigate the structures of the deeply bound kaonic atoms experimentally.
We hope that our results stimulate the experimental activities of the study of the deeply bound kaonic atoms, even it is very hard to observe these states.

\begin{acknowledgments}
We would like to thank many stimulating discussions and comments for E. Oset, H. Toki and D. Jido.
We also thank to A. Gal for useful discussions.
This work is partly supported by Grants--in--Aid for scientific research of MonbuKagakusho and Japan Society for the Promotion of Science [No. 18-8661 (H.N.) and No. 19-2831 (J.Y.)], and by CSIC and JSPS under the Japan--Spain research Cooperative Program.
The authors thank the Yukawa Institute for Theoretical Physics at Kyoto University.
 Discussions during the YKIS 2006 on `New Frontier QCD' were useful to complete this work. 

\end{acknowledgments}

\end{document}